\documentclass[12pt]{article}
\usepackage{amsmath}
\usepackage{amssymb,amsthm}
\usepackage[english]{babel}
\usepackage[textwidth=18.5cm,textheight=22cm]{geometry}
\usepackage{graphicx,epsfig}

\renewcommand{\d}{\mathrm{d}}

\newcommand{\bt}{\boldsymbol{t}}

\newcommand{\be}{\begin{equation}}
\newcommand{\ee}{\end{equation}}

\begin{document}

\title{ Regularization of Hele-Shaw flows, multiscaling expansions \\ and the Painlev\'e I equation
\thanks{Partially supported by  MEC (Ministerio de Educaci\'on y Ciencia)
project FIS2005-00319 and ESF (European Science Foundation) programme MISGAM}}
\author{L. Mart\'{\i}nez Alonso$^{1}$ and E. Medina$^{2}$
\\
\emph{$^1$ Departamento de F\'{\i}sica Te\'{o}rica II, Universidad
Complutense}\\ \emph{E28040 Madrid, Spain}\\
\emph{$^2$ Departamento de Matem\'aticas, Universidad de C\'adiz}\\
\emph{E11510 Puerto Real, C\'adiz, Spain} }
\date{} \maketitle
\maketitle

\abstract{ Critical processes of ideal integrable models of Hele-Shaw flows are considered. A regularization method based on multiscaling expansions of solutions of the KdV and Toda hierarchies characterized by string equations is proposed. Examples are exhibited in which the \emph{tritronqu\'ee} solution of the Painlev\'e-I equation turns out to provide
the leading term of the regularization}

\vspace*{1cm}

\begin{center}\begin{minipage}{12cm}

\end{minipage}
\end{center}
\newpage

\section{Introduction}

 A Hele-Shaw cell is a narrow gap between two plates filled with two fluids: say oil surrounding one or several bubbles
 of air.  Several  dispersionless (quasiclassical) limits of
 integrable systems have been found \cite{zab1}-\cite{lee3} which provide ideal models of Hele-Shaw flows in the
 absence of surface tension. Moreover, the same integrable structures also arise in random matrix models of two-dimensional quantum gravity
\cite{gin}. Integrable systems of dispersionless type are solved by
means of hodograph equations, so that generic initial conditions
reach points of \emph{gradient catastrophe} in a finite time. In the
ideal Hele-Shaw  models this feature gives rise to the cusp
formation  in the motion of the interface, while in random matrix
models it manifests itself as the critical points
 of the asymptotic expansions for large  matrix dimension $N$. In both cases a regularization mechanism of the
underlying integrable models is required.
As it is well known in random matrix theory \cite{gin}-\cite{ds3}, the \emph{double scaling limit} method provides
a regularization scheme of the large $N$ expansions which leads to models of two-dimensional quantum
gravity. On the other hand, recent work \cite{lee1}-\cite{lee3} suggests the use of methods of asymptotic solutions
of integrable systems \cite{as1}-\cite{as3} to cure the singularities in Hele-Shaw flows.

The present work is concerned with the regularization of a family of critical Hele-Shaw processes \cite{lee1}-\cite{lee3}. We mainly consider the case in which the interface develops an isolated  finger  which is close to becoming a cusp \cite{lee1}-\cite{lee2}. Then at a small scale the boundary of the finger tip  is described
by a curve
\[
Y(X)=P(X)\,\sqrt{X-v(x,t_1)},
\]
where $(X,Y)$ are Cartesian coordinates, $P(X)$ is a given
polynomial  and $v(x,t_1)$ is a particular solution of the
dispersionless KdV (Hopf) equation $ 2\,\partial_{t_1} v=3\,v\, v_x.
$ Here  $x\sim Q\,(t-t_c)$, where $t$ is the physical time, $t_c$ is
the critical time and $t_1$ is a deformation parameter. The starting
point of our analysis is the fact that $v$ satisfies the
dispersionless limit of the string equations \cite{vmo} which
characterize the one-matrix models of topological  and
two-dimensional quantum gravity \cite{wit1}-\cite{kon}. Then, from
the \emph{quasi-triviality} property \cite{dubz} of the KdV
hierarchy, there is a unique solution $u$ of the dispersionful KdV
equation \begin{equation}\label{kdve0}
\partial_{t_1} u=\dfrac{1}{4}\,\Big(\epsilon^2\, u_{xxx}+6\,u\, u_x\Big),\quad  u=\sum_{k\geq
0}\epsilon^{2k}\,u^{(k)},
\end{equation}
and its associated hierarchy such that $v=u^{(0)}$.
More precisely \cite{vmo}, the $\tau$-function
associated with $u$ is the large $N$ limit of the Kontsevich integral over hermitian matrices
\begin{align*}
\tau_N(\bt):&=\dfrac{A_N(\Theta)}{B_N(\Theta)},\quad \Theta:=\mbox{diag}(\theta_1,\ldots,\theta_N),\quad k\,t_k=\epsilon\,\mbox{Tr}(-\Theta)^k,\\\\
A_N(\Theta):&=\int \d X\,\exp \mbox{Tr}(X^2\,\Theta-\dfrac{1}{3}\,X^3),\quad
B_N(\Theta):=\int \d X\,\exp \mbox{Tr}(X^2\,\Theta).
\end{align*}
In this KdV picture, singular Hele-Shaw processes correspond to
critical points of the expansion of $u$, which in turn are
associated  to  points of \emph{gradient catastrophe} for $u^{(0)}$.
In our work  we  follow the same
 regularization procedure that is used in the one-matrix model description of two-dimensional quantum gravity \cite{ds1}-\cite{ds3}
: we apply a multiscaling limit method to obtain the leading term of the regularization of $u$ near critical points.
Then we use it to continue the Hele-Shaw flow on critical regions. We notice that according to recent work
\cite{gra1,gra2}, the multiscale regularization of KdV solutions  gives a correct asymptotic
approximation near the edges of the oscillatory zone (Whitham zone) which emerges at points of gradient catastrophe. We illustrate our strategy by regularizing
the $(2,5)$ critical finger studied in \cite{lee2}.

We also consider in this paper the critical processes of break-off and merging of Hele-Shaw bubbles \cite{lee3}. The same analysis as in the critical finger case applies, but now the ruling integrable structure is supplied by the solution of the  Toda hierarchy which underlies the large $N$ limit of the partition function of the Hermitian matrix model
\begin{equation}\label{1}
\tau_N:=\int \d H \exp\Big(\mbox{tr}(\sum_{k\geq
1}t_k\,H^k)\Big).
\end{equation}
In this case the method is illustrated with an example of regularization for the merging of two bubbles.

Our analysis makes use of the method developed by Takasaki and Takebe \cite{tt} for determining solutions of integrable systems by means of string equations  (see also \cite{mel0}-\cite{mel2}).
In the examples considered in both cases, KdV and Toda structures, the leading term of the regularization turns out
to be provided by a particular solution of the Painlev\'e-I equation (P-I)
\begin{equation}\label{pai0}
W_{\xi\xi}=6\,W^2-\,\xi,
\end{equation}
the so called \emph{tritronqu\'ee} solution discovered by P.
Boutroux \cite{bout}. This is the same function which appears in the
study of some critical processes in plasma \cite{plas} as well as in
the analysis of the critical behavior of solutions to the focusing
nonlinear Schr\"{o}dinger equation \cite{nls}. As it is known, a
different solution of P-I emerges in the random matrix models of
two-dimensional quantum gravity \cite{gin}.

\section{ Hele-Shaw flows and the KdV hierarchy}
 In the set-up considered in \cite{lee1,lee2} the cell is permeable to air but not oil. When air is injected the bubble  develops a finger whose tip is pushed away  and may become a cusp. By assuming that the finger is symmetric with respect to the $X$-axis and that the cusp is formed at the origin, then near the origin the finger turns to be described by a curve of the form
\begin{equation}\label{cur1}
Y(z):=\Big(\dfrac{\sum_{k=1}^{l+1}(k+\dfrac{1}{2})\,t_k\,z^{2k-1}}{\sqrt{z^2-v}}\Big)_\oplus\,\sqrt{z^2-v} ,\quad
X=z^2.
\end{equation}
Here the subscript $\oplus$ denotes the projection of $z$-series on
the positive powers, and
$t_k,\,k\geq 1$ are deformation parameters. The function $v$ stands for the distance between the tip and the origin and it is determined by imposing the asymptotic behaviour
\begin{equation}\label{asy}
Y(z)=\sum_{k=
1}^{l+1}(k+\dfrac{1}{2})\,t_k\,z^{2k-1}+\dfrac{x}{2\,z}+\mathcal{O}(z^{-3}),\quad
z\rightarrow\infty,
\end{equation}
where the coefficient $x$ is proportional to time $x\sim Q\,(t-t_c)$.
The resulting equation for $v$ is the hodograph equation
\begin{equation}\label{hodk}
H(\bt,v):=\sum_{k\geq 1} (2\,k+1)\, t_k \,r_k(v)+x=0,
\end{equation}
where $\bt:=(x,t_1,t_2,\ldots)$ with $t_k=0,\,k>l+1$, and $r_j$ are the coefficients of the generating function
\[
r :=\dfrac{z}{\sqrt{z^2-v}}=\sum_{k\geq
0}\dfrac{r_k(v)}{z^{2k}},\quad r_0=1.
\]

The hodograph equation \eqref{hodk} is the basic piece to solve the system of string equations
\begin{equation}\label{kdv}
(z^2)_-=0,\quad (m\,z^{-1})_-=0.
\end{equation}
for the Lax-Orlov functions of the dispersionless KP  (dKP) hierarchy
\begin{align}\label{lok}
z=p+u_0+\mathcal{O}(p^{-1}),\quad &
m=\sum_{k=1}^\infty (2\,k+1)\, t_k\, z^{2k}+x+\mathcal{O}(z^{-2}),\\
\{z,m\}&:=z_p\,m_x-z_x\,m_p=1.
\end{align}
Here and henceforth the subscripts $-$ and $+$ denote the
projection of $p$-series on the strictly negative and positive
powers, respectively. According to Theorem 1.5.1 of \cite{tt}, if
$(z,m)$ are solutions of \eqref{kdv} satisfying the asymptotic
forms \eqref{lok}, then they verify the dKP  hierarchy. More
concretely, as the first string equation means that
$
z^2=p^2+v,
$
then the function $v$ verifies the dispersionless KdV (dKdV)
hierarchy $\partial_{t_j} v=2\,\partial_x\,r_{j+1}$.

To solve the
second string equation $(m\,z^{-1})_-=0$ and satisfy the
asymptotic behaviour \eqref{lok}, one sets $m\,z^{-1}=
\sum_{k=1}^\infty (2\,k+1)\, t_k (z^{2k-1})_+$. Thus by taking
into account that $(z^{2k-1})_+=(z^{2k-2}\,r)_{\oplus}\,p$, it
follows that \eqref{kdv} reduces to \eqref{hodk}. In particular
from \eqref{cur1} and \eqref{asy} we may identify
\[
Y(z)=\dfrac{m(z)}{2\,z},
\]
so that the dynamics of the curve $Y=Y(X)$ with respect to $\bt$ is governed by the dispersionless KdV hierarchy.
The problem is that near critical points $(\bt_c,v_c)$
\begin{equation}\label{cri}
\dfrac{\partial H}{\partial v}\Big|_{\bt_c,v_c}=\ldots=\dfrac{\partial^{m-1}
H}{\partial v^{m-1}}\Big|_{\bt_c,v_c}=0,\quad \dfrac{\partial^{m} H}{\partial
v^{m}}\Big|_{\bt_c,v_c}\neq 0, \quad m\geq 2,
\end{equation}
the solutions $v$ of \eqref{hodk} are multivalued and have singular
derivatives (gradient catastroph). These situations correspond to
the critical regimes of the Hele-Shaw fingers described by
\eqref{cur1}.

In order to find  regularizations of these Hele-Shaw flows we
consider solutions of the dispersionful version of the KdV
 equation
\begin{equation}\label{kdve}
\partial_{t_1} u=\dfrac{1}{4}\,\Big(\epsilon^2\, u_{xxx}+6\,u\, u_x\Big).
\end{equation}
and the higher members of its hierarchy
$\partial_{t_j}\,u=2\partial_{x} R_{j+1}(u)$. Here $R_j$ are the
Gel'fand-Dikii polynomials determined
by\begin{equation}\label{pes3}
\epsilon^2\,(R\,R_{xx}-\dfrac{1}{2}\,R_{x}^{\,2})-2\,(z^2-v)\,R^2+2\,z^2=0,\quad
R=\sum_{n\geq 0}\dfrac{R_n(u)}{z^{2n}},
\end{equation}
or by the third-order differential equation
\begin{equation}\label{third}
\partial_{x}\,R_{n+1}=\Big(\dfrac{1}{4}\epsilon^2\,\partial_{x}^3+v\,\partial_x+\dfrac{1}{2}\,u_{x}\Big)\,R_n.
\end{equation}

Our first observation is that there is solvable dispersionful
version of the string equations \eqref{kdv} given by
\begin{equation}\label{kdvq}
(L^2)_-=0,\quad \dfrac{1}{2}(M\,L^{-1}-\dfrac{\epsilon}{2}
\,L^{-2})_-=0.
\end{equation}
Here $L$ and $M$ are Lax-Orlov operators ($[L,M]=\epsilon$)
\begin{equation}\label{sati30}
 L=\epsilon\,\partial_{x}+u_1\,\partial_{x}^{-1}+\cdots,
\quad M=\sum_{j\geq 1} (2\,j+1)\,t_j
\,L^{2\,j}+t_0+\mathcal{O}(L^{-2}),
\end{equation}
of the dispersionful KP hierarchy. Now, the $\pm$ parts of a
pseudo-differential operator denote  the truncations of
$\partial_{x}$-series in the positive and strictly negative power
terms, respectively. According to Proposition 1.7.11 of \cite{tt},
given a solution $(L,M)$ of \eqref{kdvq} satisfying \eqref{sati30}
and $[L,M]=\epsilon$, then they are Lax-Orlov operators of the
dispersionful KP hierarchy.
The first string equation in \eqref{kdvq} constitutes the KdV reduction condition and leads to a Lax operator of the form
\[
L=(\epsilon^2\,\partial_x^2+u)^{\frac{1}{2}}.
\]
The second string equation together with the asymptotic condition
on $M$ and  $[L,M]=\epsilon$ can be satisfied by setting
\[
M\,L^{-1}-\dfrac{\epsilon}{2}\,L^{-2}= \sum_{j=1}^\infty
(2j+1)\,t_j\,(L^{2j-1})_+.
\]
Thus, by taking into account the identity
$
[L^2,(L^{2j+1})_+]=-2\epsilon\,\partial_x\,R_{j+1},
$
the problem reduces to finding solutions of the form
\begin{equation}\label{exp}
 u=\sum_{k\geq
0}\epsilon^{2k}\,u^{(k)},
\end{equation}
verifying
\begin{equation}\label{4.5}
\sum_{k\geq 1}(2k+1)\,t_k
R_{k}+x=0,
\end{equation}
or equivalently
\begin{equation}\label{hoin222}
\oint_{\gamma}\dfrac{d z}{2\pi i\,z}\, V_z\,R\,+x =0,
\end{equation}
where $V$ denotes the function $ V:=\sum_{j\geq 1} z^{2j+1}\,t_j $
and $\gamma$ is a large positively oriented closed path.

To solve this equation for $u$ one may use \eqref{pes3} to determine the coefficients of the $\epsilon$-expansion
\begin{equation}\label{rep}
R=\sum_{n\geq 0}\epsilon^{2n}\,R^{(n)},\quad R^{(0)}\,=\,\frac{z}{(z^2-u^{(0)})^{\frac{1}{2}}},
\end{equation}
and  substitute them in \eqref{hoin222} to get the system \vspace{0.3cm}
\begin{equation}\everymath{\displaystyle}\label{hoin4}
\oint_{\gamma}\dfrac{d z}{2\pi i\,z}\,
V_z(z)\,R^{(l)}(z)
+\delta_{l0}\,x=0.
\end{equation}

From \eqref{pes3}, \eqref{rep} and \eqref{hoin4}  an iterative scheme for obtaining the expansion \eqref{exp} of $u$
 follows. In particular,  for $l=0$ it reduces to the hodograph equation \eqref{hodk} for $u^{(0)}$. This means that the expansion
 \eqref{exp} is not valid near critical points \eqref{cri} so that a different expansion must be used to
 construct a  regularization of the solutions of \eqref{hodk} on critical regions.

\section{Multiscaling expansions and asymptotic matching }

Given  a $m$-th order critical point $(\bt_c,v_c)$   \eqref{cri}  of \eqref{hodk}, let us introduce a new small parameter
$\tilde{\epsilon}$ and new variables $\tilde{t}_j$ and $\tilde{x} $ given by
\begin{equation}\label{dl}
\tilde{\epsilon}:=\epsilon^{\frac{2}{2m+1}},\quad
t_j=t_{c,j}+\tilde{\epsilon}^m\,\tilde{t}_j,\quad
x=x_c+\tilde{\epsilon}^m\,\tilde{x}.
\end{equation}
 Let us now look for solutions to \eqref{hoin222} of the form
\begin{equation}\label{dl1}
 u=v_c+\sum_{n\geq
1}\tilde{\epsilon}^{n}\,\tilde{u}^{(n)},
\end{equation}
where now $R$ is expanded in the form $R=\sum_{n\geq
0}\tilde{\epsilon}^{n}\,\widetilde{R}^{(n)}$.
To determine the coefficients of $R$, we first observe that $\epsilon\,\partial_{x}=\tilde{\epsilon}^{1/2}\,\partial_{\tilde{x}}$,
so that \eqref{pes3} can be rewritten as
\begin{equation}\label{pes2a}
\tilde{\epsilon}\,(R\,R_{\tilde{x}\tilde{x}}-\dfrac{1}{2}\,R_{\tilde{x}}^2)-2\,(z^2-u)\,R^2+2\,z^2=0.
\end{equation}
From this equation and taking into account that $v_{c,\tilde{x}}\equiv 0$ one deduces a recursion relation
for the coefficients $\widetilde{R}^{(n)}$ and that they can be expressed in the form
\begin{equation}\label{RT01}
\widetilde{R}^{(n)}=\widetilde{R}^{(0)}\,
\sum_{r=1}^{n}\dfrac{\widetilde{G}_{n,r}}{(z^2-v_c)
^r}, \quad\widetilde{R}^{(0)}\,=\,\frac{z}{(z^2-v_c)^{\frac{1}{2}}},
\ee
where the functions $\widetilde{G}_{n,r}$  are differential polynomials in
$\tilde{u}^{(k)},\,0\leq k\leq n-r+1$. In particular, from \eqref{pes2a} it follows that the leading coefficients $G_n:=\widetilde{G}_{n,n},\quad G_0=1$ satisfy
the same recursion relation as that arising from \eqref{pes3}  (with $\epsilon\equiv 1$)
for the Gel'fand-Dikii polynomials $R_n$. In other words
\[
G_n=R_n(\tilde{u}^{(1)}).
\]

\vspace{0.2cm}

If we now substitute \eqref{dl}-\eqref{dl1} in \eqref{hoin222} and identify coefficients of $\tilde{\epsilon}$-powers we get
the system of equations
\begin{equation}\label{sysdl}\everymath{\displaystyle}
\begin{cases}
\oint_{\gamma}\dfrac{d z}{2\pi i\,z} \, V_z(\bt_c)\,\widetilde{R}^{(k)}+\delta_{k0}\,x_c=0,\quad k=0,\ldots,m-1
,\\\\
\oint_{\gamma}\dfrac{d z}{2\pi i\,z} \, V_z(\bt_c)\,\widetilde{R}^{(k)}+
\oint_{\gamma}\dfrac{d z}{2\pi i\,z} \, V_z(\tilde{\bt})\,\widetilde{R}^{(k-m)}+
\delta_{km}\,\tilde{x}=0,
\quad k\geq m.
\end{cases}
\end{equation}
Since $v_c$ is a $m$-th order critical point of \eqref{cri} we have
that
\[
\oint_{\gamma}dz\,\dfrac{V_z(\bt_c)}{(z^2-v_c)
^{\frac{2k+1}{2}}}+\delta_{k0}\,x_c
=0,\quad k=0,\ldots,m-1.
\]
Hence, in view of \eqref{RT01}  the first $m$ equations of
the system  are identically satisfied, while the remaining ones
determine recursively
the coefficients $\tilde{u}^{(n)}$ for $n\geq 1$.   In particular
for $k=m$ we get the following differential equation for the leading
contribution $\tilde{u}^{(1)}$
\begin{equation}\label{doug}
\Big(\sum_{j\geq 1}c_{jm}(v_c)\,t_{c,j}\Big)\,
R_m(\tilde{u}^{(1)})+\sum_{j\geq 1}c_{j0}(v_c)\,\tilde{t}_j+\tilde{x}=0,
\end{equation}
where $R_m(\tilde{u}^{(1)})$ is the $m$-th Gelfand-Dikii polynomial in $\tilde{u}^{(1)}$ and
\[
c_{jr}(v_c)=(2j+1)\oint_{\gamma}\dfrac{dz}{2\pi i}\,\dfrac{z^{2j}}{(z^2-v_c)
^{\frac{2r+1}{2}}}
\]
From \eqref{doug} it follows that $\tilde{v}^{(1)}$ depends on the rescaled variables $\tilde{\bt}$ through the linear combination
$
\sum_{j\geq 1}c_{j0}(v_c)\,\tilde{t}_j+\tilde{x},
$
so that
$
\partial_{\tilde{t}_j}\,\tilde{u}^{(1)}=c_{j0}(v_c)\,
\tilde{u}^{(1)}_{\tilde{x}}$.

\vspace{0.3cm}

Having in mind the applications to Hele-Shaw flows, we must match
the solutions \eqref{exp} and \eqref{dl1} for $t_k=t_{c,k},\,
\forall k\geq 1$ and $x$ varying  on some overlap interval such us
as $ \epsilon\rightarrow 0^+$ we have $ x-x_c\rightarrow 0^-$ and
$\tilde{x}\rightarrow -\infty$. To this end we observe that since
${u}^{(0)}$ satisfies the hodograph equation \eqref{hodk}, then near
an $m$-th order critical point $(\bt_c,v_c)$ it behaves as
\begin{equation}\label{in}
{u}^{(0)}(x,t_{c,1},t_{c,2},\ldots)\sim v_c+\sqrt[m]{c\,(x-x_c)},\quad
c:=-m!\,\Big(\dfrac{\partial^{m} H}{\partial
u^{m}}(\bt_c,v_c)\Big)^{-1}.
\end{equation}
Hence,  the solutions $u={u}^{(0)}+\mathcal{O}(\epsilon)$ and
$
u= v_c+\tilde{\epsilon}\,\tilde{u}^{(1)}+\mathcal{O}(\tilde{\epsilon}^2)
$
match to first order in $\tilde{\epsilon}$ provided $\tilde{u}^{(1)}$ is a solution of the differential equation
\eqref{doug} such that
\begin{equation}\label{au1}
\tilde{u}^{(1)}\sim \sqrt[m]{c\,\tilde{x}}, \quad \tilde{x}\rightarrow -\infty.
\end{equation}

Let us  illustrate our analysis by studying  the (2,5) critical finger
considered in \cite{lee2}. If we set  $t_j=0$ for all $j\geq
2$ except $t_3=2/7$, the hodograph equation
\eqref{hodk} for $u={u}^{(0)}$ becomes
\begin{equation}\label{fc}
\dfrac{5}{8}\,u^3+\dfrac{3}{2}\,t_1\,u+x=0,
\end{equation}
so that we have
\begin{equation}\label{u0}
{u}^{(0)}\,=\,\left(\frac{2}{5}\right)^{2/3}
\left(\sqrt{5(4{t_1}^3+5 x^2)}-5 x\right)^{\frac{1}{3}}-
\frac{2 \sqrt[3]{\frac{2}{5}}\,{t_1}}{\left(\sqrt{5(4 {t_1}^3+5 x^2)}-5x\right)^{\frac{1}{3}}}.
\end{equation}
We consider the case $t_1<0$ which leads to a cusp of the type  $Y^2\sim X^3$ which can not be continued \cite{lee2}. The hodograph equation \eqref{fc} has a $2$-nd order critical point at
\[
x_c\,=\,-t_{c,1}\sqrt{-\frac{4}{5}t_{c,1}},\quad
v_c=\sqrt{-\frac{4}{5}t_{c,1}},
\]
so that the solution behaves as
\[
u\sim v_c+\sqrt{-\dfrac{8}{15\,v_c}\,(x-x_c)},\quad x\rightarrow
x_c^-.
\]
If we take $t_1=t_{1,c}$ (i.e. $\tilde{t}_1=0$) equation
\eqref{doug} becomes
\begin{equation}\label{out}
\tilde{u}_{\tilde{x}\tilde{x}}^{(1)}\,+\,3\,\tilde{u}^{(1)^2}\,=\,-\frac{8}{5\,v_c}\tilde{x},
\end{equation}
and the  matching condition
requires
\begin{equation}\label{asym}
\tilde{u}^{(1)}\sim\sqrt{-\frac{8}{15\,v_c}\tilde{x}}.
\end{equation}
By introducing the rescalings
\[
W:=-\dfrac{1}{2}\Big(\dfrac{5\,v_c}{4}\Big)^{2/5}\,\tilde{u}^{(1)},\quad
\xi:=-\Big(\dfrac{4}{5\,v_c}\Big)^{1/5}\,\tilde{x},
\]
we have that the differential equation \eqref{out}  reduces to the
Painlev\'e I (P-I) equation \eqref{pai0},  while the matching
condition reads
\begin{equation}\label{machp}
W\sim -\sqrt{\xi/6},\quad \xi\rightarrow \infty.
\end{equation}
 As
it is known \cite{kap}, the tritronqu\'ee solution discovered by P. Boutroux \cite{bout} is the unique solution of P-I having no poles in the sector $|\mbox{arg}\, \xi|<4\pi/5$ for sufficiently large $|\xi|$. Moreover, it has no poles on the positive real axis \cite{josi}.

One may now proceed \cite{nls,josi} by taking a numerical
approximation to the tritronqu\'ee solution for large positive
values of $\xi$ and use it to supply initial data for the P-I
equation. To this end we have taken the approximation provided by Eq.(3.3)
of \cite{kap}, and have set $t_1=t_{c,1}=-\frac{4}{5}$ so that
$x_c=\frac{16}{25}=0.64$ and $u_c=\frac{4}{5}$. Thus, with the aid
of the ODE solver of Mathematica we construct a numerical solution
that matches with the dispersionless approximation \eqref{u0}. The
numerical analysis shows that for $\epsilon=10^{-5}$ we have that
$|u^{(0)}-(v_c+\tilde{\epsilon}\tilde{u}^{(1)})|<5\cdot 10^{-4}$ in
the $x$-interval $(0.6365,0.6395)$. As both functions take values
larger than 0.8, the relative error on this interval is smaller than
0.000625. The matching between both solutions on the
$x$-intervals $[0.58,0.64022]$ and $[0.6,0.6403]$ can be observed in
Fig.1.

\begin{center}
\begin{figure}[!ht]
\centering
\includegraphics[width=10cm]{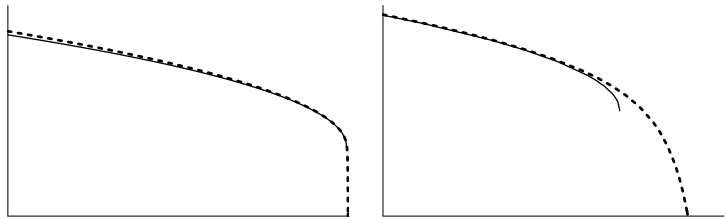}
\caption{ $u^{(0)}$ (solid line) and
$u_c+\tilde{\epsilon}\tilde{u}^{(1)}$ (dashed line) }
\end{figure}
\end{center}

In this way, near the critical point the regularized Hele-Shaw dynamics associated to our asymptotic
approximation  is given by the curve
\begin{equation}\label{curve}
Y(X)\,=\,\left(X^2+\frac{u}{2}X+\frac{3}{8}u^2-\frac{6}{5}\right)\sqrt{X-u}.
\end{equation}
 The function $u$ is given by $u^{(0)}(x)$ for $x<0.638$,  and by
$v_c+\tilde{\epsilon}\tilde{u}^{(1)}$ for $x\geq0.638$. It is
defined until a certain $x^*$ which corresponds to the first pole of
the tritronqu\'ee solution on the negative $\xi$-axis. Figures 2
and 3 exhibit  a sequence  between $x=0.6$ and $x=0.6402302$. Notice that  $u$ is a decreasing function of $x$ and that
a cusp is formed when one of the roots of the polynomial
$P(X)=X^2+\frac{u}{2}X+\frac{3}{8}u^2-\frac{6}{5}$ coalesces with $u$. Thus, for $x<0.64$ the
 tip of the finger moves to the left,  and it starts forming a cusp  as $x$ is close to a certain value slightly higher
 that $0.64$ ($u$ coincides with
the largest root of $P$). Next a  bubble appears at the tip of the finger. Subsequently a new cusp
forms in this bubble
($u$ coincides with the smallest root of $P$), and  a second bubble grows from this cusp
while the first bubble declines until it annihilates. Finally, the remaining bubble is  absorbed by the
finger (when both roots of $P$ coincide).

\begin{center}
\begin{figure}[!ht]
\centering
\includegraphics[width=14cm]{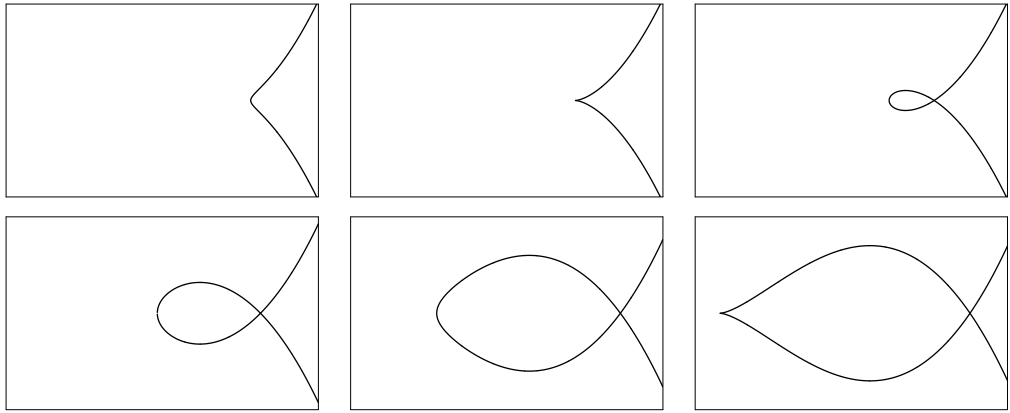}
\caption{Cusp formation and creation of a bubble which develops a new
cusp}
\end{figure}
\end{center}

\begin{center}
\begin{figure}[!ht]
\centering
\includegraphics[width=14cm]{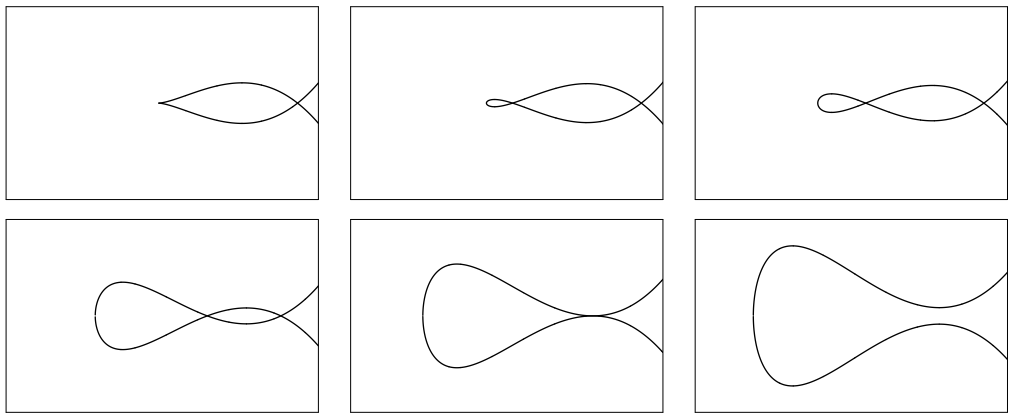}
\caption{Emergence, annihilation and absorbtion of  bubbles}
\end{figure}
\end{center}

\subsection{Hele-Shaw flows and the Toda hierarchy}

In  the Hele-Shaw set-up considered in \cite{lee3} air is injected in two fixed points of a simply-connected air bubble making
 the bubble  break into two emergent bubbles.  Before the break-off the interface oil-air
 remains free of cusp-like singularities and develops a smooth neck. The reversed evolution describes the merging of two bubbles.

The analysis of \cite{lee3} concludes that after the break-off the
local structure of a small part of the interface containing the tips
of the bubbles falls into universal classes characterized by two
even integers $(4\,n, 2),\, n\geq 1,$ and a finite number $2n$ of
real deformation parameters $t_k$. By assuming symmetry of the curve with
respect to the $X$-axis, the general solution for the curve and the
potential in the $(4\,n, 2)$ class are
\begin{equation}\label{den}
Y(z):=\Big(\dfrac{\sum_{k=1}^{2n} (k+1)\,t_{k+1}\,z^k}{\sqrt{(z-a)(z-b)}}\Big)_\oplus\,\sqrt{(z-a)(z-b)},\quad X=z.
\end{equation}
where $a$ and $b$ are the positions of the bubbles tips.
 Due to the physical assumptions of the problem, the expansion
\begin{equation}\label{hoh}
Y(z)=\sum_{k=1}^{2n}
(k+1)\,t_{k+1}\,z^{k}+\sum_{k=0}^{\infty}\dfrac{Y_n}{z^n},\quad
z\rightarrow\infty,
\end{equation}
of the function $Y$ must satisfy two conditions $Y_0=t$ (physical time) and $Y_1=0$ which determine the positions $a$, $b$ of the tips.
As it was shown in \cite{lee3}, imposing these two conditions on \eqref{hoh} leads to a pair of  hodograph equations
\begin{equation}\label{ho}
\sum_{k=1}^\infty k\,t_k
r_{k-1}(u,v)=0,\quad
\sum_{k=1}^\infty
k\,t_k\,r_k(u,v)+2\,x=0,
\end{equation}
where $t_k=0$ for $k>2n$ and
\begin{align}\label{d7b}
\nonumber r :&=\dfrac{z}{\sqrt{(z-u)^2-4v}}=\sum_{k\geq
0}\dfrac{r_k(u,v)}{z^k},\\\\
\nonumber
a:&=u-2\,\sqrt{v},\quad b=u+2\,\sqrt{v}.
\end{align}
These equations arise in the dispersionless AKNS hierarcy \cite{lee3} . However, it is straightforward to see
\cite{mel0} that by setting
\begin{equation}\label{sett}
 Y=2\,m-\sum_{k=1}^{2n} (k+1)\,t_{k+1}\,z^k,\quad x=0,\quad t=t_1,\quad t_j=0,\quad  \forall j\geq 2n+2,
\end{equation}
they also coincide with the hodograph equations  which solve the  string equations
\begin{equation}\label{11}
\bar{z}=z,\quad \overline{m}=m,
\end{equation}
for the Lax-Orlov functions
\begin{align*}
z=&p+u_0+\mathcal{O}(p^{-1}),\quad m=\sum_{k=1}^\infty k\, t_k z^{k-1}+\dfrac{x}{z}+ \mathcal{O}(z^{-2}),\\
\bar{z}=&\dfrac{v_0}{p}+v_1+\mathcal{O}(p),\quad \overline{m}=\sum_{=k1}^\infty
j\,\bar{t}_k \bar{z}^{k-1}-\dfrac{x}{\bar{z}}+\mathcal{O}(\bar{z}^{-2}),
\end{align*}
of the dispersionless 2-Toda (d2-Toda)   hierarchy.
In particular the first string equation represents the 1-Toda reduction $z=\bar{z}=p+u+v\,p^{-1}$.

In order to regularize the critical points of the hodograph equations \eqref{ho} one may consider appropriate solutions of the dispersionful version of the Toda hierarchy . The natural candidates are provided by the string equations
\begin{equation}\label{sto}
L=\bar{L},\quad M=\overline{M},
\end{equation}
for the Lax-Orlov operators $[L,M]=[\bar{L},\overline{M}]=\epsilon$
of the dispersionful 2-Toda hierarchy \cite{mel1,mel2}. The first equation determines the 1-Toda reduction  $L=\bar{L}=\Lambda+u+v\,\Lambda^{-1},\,(\Lambda:=\exp{(\epsilon\,\partial_x)})$. The system \eqref{sto} characterizes the partition function of the hermitian matrix model in the large-$N$ limit . In this way the well-known double-scaling limit method for this matrix model can be used to regularize the critical points of \eqref{ho}.

As an example let us analyze the critical process of a merging of two bubbles studied in
section VII of \cite{lee3} . Thus we set $t_2=0$, $t_n=0$, $n>3$,
$t:=t_1$ so that \eqref{ho} reduces to
\begin{equation}\label{hod0}
t+3t_3\,(u^2+2v)\,=\,0,\quad
6t_3v\,u+x\,=\,0.
\end{equation}
There is a 2nd-order critical point:
$v_c\,=\,u_c^2,$
with $u_c$ satisfying
\begin{equation}\label{crit}
t_c+9t_3u_c^2\,=\,0,\quad
6t_3u_c^3+x_c\,=\,0,
\end{equation}
and consequently
$$4t_c^3+81t_3x_c^2\,=\,0.$$

On the other hand, the system of string equations \eqref{sto} of the dispersionful 2-Toda reduces to
\begin{equation}\label{hod1}
t+3t_3(u^2+v+v(x+\epsilon))\,=\,0,\quad
3t_3\,(u+u(x-\epsilon))\,v+x\,=\,0.
\end{equation}
where  $(u,v)$ are characterized by  expansions of the form
\begin{equation}\label{semc}
u=\sum_{k\geq 0}\epsilon^k\,u^{(k)},\quad v=\sum_{k\geq
0}\epsilon^{2k}\,v^{(2k)}.
\end{equation}
Obviously, from \eqref{hod1} it follows that the leading terms
$(u^{(0)},v^{(0)})$ satisfy the hodograph equations \eqref{ho}.
To deal with  solutions of \eqref{hod1} near critical points
$(\bt_c,u^{(0)},v^{(0)})=(\bt_c,u_c,v_c)$ of \eqref{ho},
 one introduces
\begin{equation}\label{dl2}
\tilde{\epsilon}:=\epsilon^{\frac{1}{5}},\quad
x\,=\,x_c+\tilde{\epsilon}^{4}\tilde{x},\quad
t\,=\,t_c+\tilde{\epsilon}^{4}\tilde{t}.
\end{equation}
It can be proved that \eqref{hod1} admits solutions of the form
\be\label{p1}
u\,=\,u_c+\sum_{j=2}^{\infty} \tilde{\epsilon}^jU^{(j)},\quad
v=\,v_c+\sum_{j=1}^{\infty}\tilde{\epsilon}^{2j} V^{(2j)}.
\ee

Thus, by equating  the coefficients of  $\tilde{\epsilon}^j$ for $j=2,3$ and $4$ in \eqref{hod1}
we get
\begin{align}\label{sist}
\nonumber
U^{(2)}\,&=\,-\frac{1}{u_c}V^{(2)},\quad
U^{(3)}\,=\,-\frac{1}{2u_c}V^{(2)}_{\tilde{x}},\\
2(V^{(4)}&+u_cU^{(4)})\,=\,-\frac{\tilde{t}}{3t_3}-(U^{(2)})^2-\frac{1}{2}\,V^{(2)}_{\tilde{x}\tilde{x}},\\
\nonumber
\frac{1}{2}\partial_{\tilde{x}}^2&(V^{(2)}-u_cU^{(2)})+u_cU^{(3)}_{\tilde{x}}-\frac{2}{u_c}U^{(2)}V^{(2)}
+(U^{(2)})^2\,=\,\frac{1}{3t_3u_c}(\tilde{x}-u_c\tilde{t}).
\end{align}
This provides us
with an expression of $U^{(4)}$ in terms of $(V^{(2)},V^{(4)})$ and implies that $V^{(2)}$ verifies
\begin{equation}\label{eqV2}
V^{(2)}_{\tilde{x}\tilde{x}}+\frac{6}{u_c^2}(V^{(2)})^2\,=\,\frac{2}{3t_3u_c}(\tilde{x}-u_c\tilde{t}).
\end{equation}

Near the critical point the   solution of \eqref{hod0} behaves  as
\[
u\,\sim\,u_c\,-\,\frac{1}{3}\sqrt{\frac{1}{t_3}(t_c-t)},\quad
v\,\sim\,v_c\,+\,\frac{u_c}{3}\sqrt{\frac{1}{t_3}(t_c-t)} \quad
\mbox{as} \quad t\rightarrow t_c^-
\]
so that matching requires a solution of \eqref{eqV2} satisfying
\begin{equation}\label{asym2}
V^{(2)}\,\sim\,\frac{u_c}{3}\sqrt{-\frac{\tilde{t}}{t_3}},\quad
\mbox{as} \quad \tilde{t}\rightarrow-\infty.
\end{equation}
Now, if we set $x=x_c=1$ and introduce the change of variables
\[
W\,=\,-\left(\frac{2u_c^2}{3t_3}\right)^{-2/5}V^{(2)},\quad
\xi\,=-\,\left(\frac{2u_c^2}{3t_3}\right)^{1/5}\tilde{t},
\]
it follows that $W$ must satisfy the P-I equation \eqref{pai0} and the asymptotic condition
$W\sim-\sqrt{\frac{\xi}{6}}$ as $\xi\rightarrow \infty$, so that it must be the tritronqu\'ee solution of P-I.

Thus near the critical point the regularized Hele-Shaw dynamics of this example is characterized by the curve
\[
Y(X)=3t_3(X+u)\sqrt{(X-u)^2-4v},
\]
where
\[
u=u_c-\frac{\tilde{\epsilon}^2}{u_c}V^{(2)},\quad
v=v_c+\tilde{\epsilon}^2V^{(2)},
\]
The resulting process represents the merging of the tips of two bubbles. It turns out that the right bubble  develops a cusp, then a new bubble appears at this cusp and it grows until it merges with the tip of the left bubble.

\vspace{0.3cm}

\noindent {\large{\bf Acknowledgements}}

\vspace{0.3cm}
The authors  wish to thank the  Spanish Ministerio de Educaci\'on y Ciencia (research project FIS2005-00319) and
 the European Science Foundation (MISGAM programme) for their  support.

\end{document}